\documentclass[twocolumn]{aastex631}

\usepackage{amsmath}
\shorttitle{Turbulent magneto-genesis}
\shortauthors{Pucci et al.}
\graphicspath{{./}{figures/}}

\begin{document}

\title{Turbulent magneto-genesis in a collisionless plasma}

\author{F. Pucci}
\email{francesco.pucci@kuleuven.be}
\affiliation{Centre for Mathematical Plasma Astrophysics, Department of Mathematics, KU Leuven, Celestijnenlaan 200B, 3001 Leuven, Belgium}
\affiliation{Istituto per la Scienza e Teconologia dei Plasmi, Consiglio Nazionale delle Ricerche (ISTP-CNR), Via Amendola 122/D, 70126 Bari, Italy}

\author{M. Viviani}
\affiliation{Dipartimento di Fisica, Universit\`a della Calabria, I-87036 Cosenza, Italy}

\author{F. Valentini}
\affiliation{Dipartimento di Fisica, Universit\`a della Calabria, I-87036 Cosenza, Italy}

\author{G. Lapenta}
\affiliation{Centre for Mathematical Plasma Astrophysics, Department of Mathematics, KU Leuven, Celestijnenlaan 200B, 3001 Leuven, Belgium}

\author{W. H. Matthaeus}
\affiliation{Department of Physics and Astronomy, University of Delaware, Newark, DE 19716, USA}

\author{S. Servidio}
\affiliation{Dipartimento di Fisica, Universit\`a della Calabria, I-87036 Cosenza, Italy}



\begin{abstract}

We investigate an efficient mechanism 
for generating magnetic fields in turbulent, collisionless plasmas. By using fully kinetic, particle-in-cell simulations of an initially non-magnetized plasma, we inspect the genesis of magnetization, in a nonlinear regime. The complex motion is initiated via a Taylor-Green vortex, and the plasma locally develops strong electron temperature anisotropy, due to the strain tensor of the turbulent flow. Subsequently, in a domino effect, the anisotropy triggers a Weibel instability, localized in space. In such active wave-particle interaction regions, the magnetic field seed grows exponentially and spreads to larger scales due to the interaction with the underlying stirring motion. Such a self-feeding process might explain magneto-genesis in a variety of astrophysical plasmas, wherever turbulence is present.

\end{abstract}



\section{Introduction} \label{sec:intro}

Magnetic fields permeate the Universe, but their origin still represents an open question \cite{KZ2008}, despite years of studies (see \cite{Widrow2012} and references therein).  The magnetic fields of most stars and galaxies are believed to be sustained and amplified by hydromagnetic dynamo action \cite{Parker70}, an essential element of which is turbulence \cite{KraichnanNagarajan67,Pouquet76,mininni2003dynamo,Br2018}. In turn, turbulent motion is stirred by the evolution of baryonic and dark matter, with stellar collisions being suggested as a possible mechanism for increasing local shearing motions~\cite{colgate2001origin}.
For extragalactic plasmas, also dynamos in non-collisional and weakly collisional plasmas have been investigated \cite{RCSV2016,StOK2018,Rincon2019,PJBTHFSA2020,StOKSS2020}. In both these scenarios, an initial seed of magnetic field is assumed. It is therefore essential to investigate the origin of such seeds from which magnetic fields can emerge and grow as we can observe them today on large and small scales.

The generation of seed magnetic fields is relevant not only in astrophysics but also in laboratory settings such as laser-plasma experiments \cite{SLFS16}. A proposed explanation invokes effects beyond standard magnetohydrodynamics, such as the Biermann battery \cite{Biermann50}, which relies on the different inertial response of electrons and ions to a pressure gradient. 

In the case of weakly-collisional media,  one of the most efficient candidates for the so-called magneto-genesis is the kinetic Weibel instability \cite{Weibel59}. This process, based on the instability of strongly anisotropic particle distribution functions, has been verified in a variety of simulations and plasma experiments \cite{huntington2015observation,SS2020}. In the collisionless case, it dominates the dynamics at small scales at $\ell \sim d_e$, where $d_e$ is the electron inertial length. Although such kinetic instability is a very powerful magnetic field source, it requires an existing highly non-Maxwellian plasma, namely an ad-hoc, unstable velocity distribution. In the Weibel scenario, such anisotropic distribution is a given ingredient.

More recently, in a sequence of inspiring works, it has been suggested that collisionless plasmas might develop large temperature anisotropy, and hence non-Maxwellian distributions, not only via an existing magnetic field but also via gradients of fluid-like variables. The temperature anisotropy can be generated via shearing motions in which gradients of the particle bulk velocities are present \cite{CPCDJ2014,DPC2016}. In principle, the production of these anisotropies can make the plasma Weibel-unstable, with subsequent production of a small-scale magnetic field produced by 
growth and nonlinear saturation of the instability. However, this possibility hasn't been explored yet.

In this Letter, we establish a connection between the above elements and investigate the possibility that plasma turbulence provides locally strong velocity gradients, initiating the magneto-genesis. 
We inspect the generation of a magnetic field in an initially isotropic, Maxwellian plasma, with ions and electrons, via full-Vlasov simulations. We perturb such a collisionless system via a vortical, turbulent motion. 
In the turbulent field, local shears initiate electron pressure anisotropy, which subsequently drives the Weibel instability. In this chain reaction, the magnetic field is then amplified due to the kinetic plasma interaction with the turbulent, stirring flow. 

\section{Method} \label{sec:style}

We solve numerically the Vlasov-Maxwell set of equations for a  plasma made of of ions ($i$) and electrons ($e$), by using the fully kinetic semi-implicit particle in cell code iPic3D \cite{MLR2010}. 
The computational domain is a cubic box of length $L=20d_i$ and the number of cells is $512^3$. We use a  reduced ion to electron mass ratio $m_i/m_e=256$, resulting in a grid  size $\Delta_{xyz}\simeq0.6d_e\simeq18\lambda_D$, where $d_e$ is the  electron skin-depth and $\lambda_D$ is the Debye length. For the time  step $\Delta t$, we chose  $\Delta t=0.0625\omega_{pi}^{-1}=\omega_{pe}^{-1}$,  where $\omega_{pi}$ and $\omega_{pe}$ are the ion and electron  plasma frequency, respectively. The initial velocity field is prescribed in the form of a Taylor-Green vortex \citep{TG37}, with the bulk flow of particles described by
\begin{align}\label{eq:TG}
    {\bf u}_s \left( x,y,z \right) = 
    u_{s0} \left[ {\rm sin} (\kappa_0x) \,{\rm cos} (\kappa_0y) \,{\rm cos} (\kappa_0z) \,\hat{\bf e}_x \right.\nonumber\\
    - \left.{\rm cos} (\kappa_0x) \, {\rm sin} (\kappa_0y) \,{\rm cos} (\kappa_0z) \, \hat{\bf e}_y \right],
\end{align}
where $s= i, e$ indicates the particles species, $u_{i0} = u_{e0} = 0.03c$ is the large-scale flow, $c$ is the speed of light and $\kappa_0=2\pi/L$.  At the beginning of the simulation, the electric and magnetic field are zero and the distributions of electrons and ions are Maxwellians  with thermal speed $v_{th,e}=0.035c$ and $v_{th,i}=0.005c$, respectively. The density is uniform and the net charge is zero. We populate each cell with $500$ particles. 
Periodic boundary conditions are used in all three Cartesian directions.
Here we show results for the highest resolution run, although we performed a convergence study by varying the mesh resolution and the number of particles per cell.

\section{Results} \label{sec:floats}

\begin{figure}
\includegraphics[width=\columnwidth]{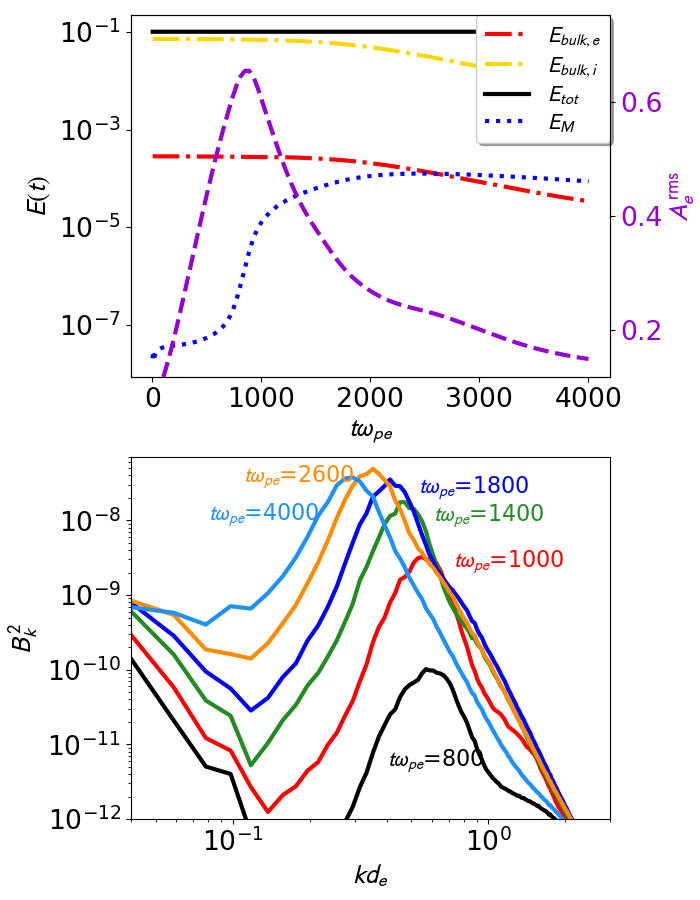}
\caption{ (Top) Time evolution of the total (black), ion and electron bulk-kinetic (yellow and red), and magnetic (dot-blue) energies. Electron temperature anisotropy is represented with (purple) dashed line, labelled on the right axis. (Bottom) Magnetic energy spectra at different times.}
\label{fig:enet}
\end{figure}

The flow quickly develops complexity, producing small-scale eddies, thus establishing a fully turbulent cascade \cite{Olshevsky2018}. The upper panel in Fig.~\ref{fig:enet} shows the time evolution of some relevant energy components, together with the total energy, which is well-conserved. Beginning with the state described in Eq.~(\ref{eq:TG}),
the bulk kinetic energy of the flow decays, as expected in turbulence, producing smaller structures and perturbing the other fields. After an initial transient, the magnetic energy starts an exponential growth at $t\sim800\omega_{pe}^{-1}$. 

To determine the characteristic spatial scale of the emerging magnetic field, we computed the isotropic magnetic energy spectra, as a function of $k$, at different times of the simulation. As it can be seen from Fig.~\ref{fig:enet} (bottom), the energy starts growing at a fixed mode  $k\simeq 0.6d_e^{-1}$, until $t=1000\omega_{pe}^{-1}$ \cite{SLFS16}. After this fast growth phase, the magnetic energy increase is slower and is back-transferred in $k$, towards larger scales. 

\begin{figure}
\includegraphics[width=\columnwidth]{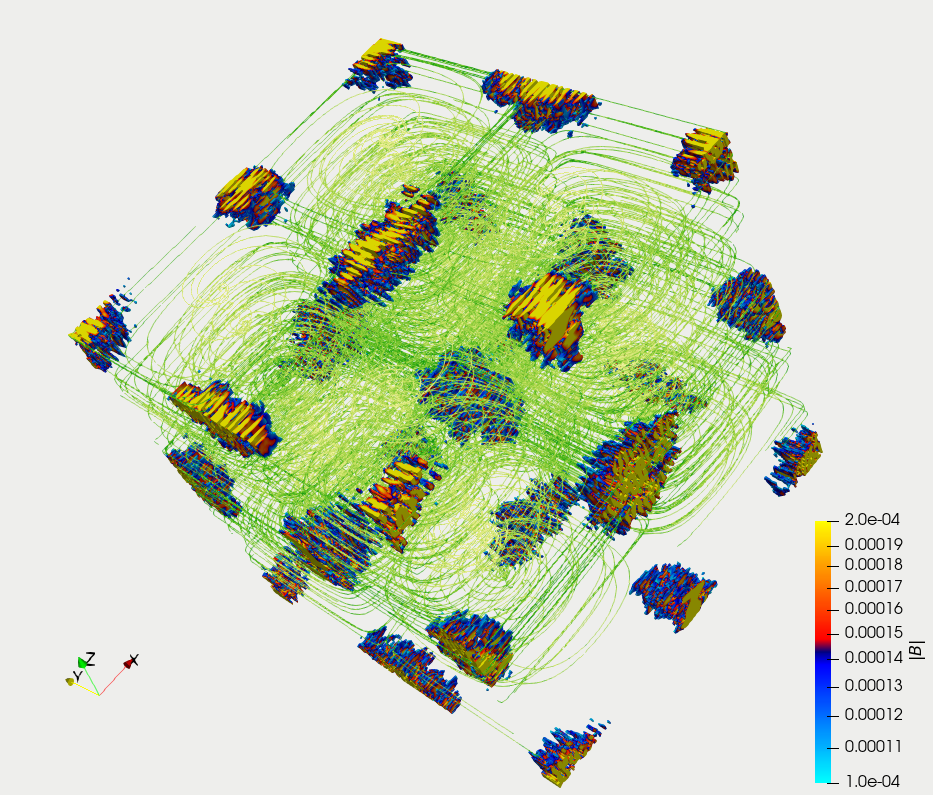}
\caption{Rendering of the Taylor Green turbulent field, with filled contours of the magnetic field $|\bf B|$ (colored patches) and streamlines of ions velocity (green), at $t=800\omega_{pe}^{-1}$.}\label{fig:3Dplots}
\end{figure}

In order to understand the origin of $B$, we analyze the structure of the electron velocity distribution functions, in absence of preferred directions \cite{SVCF2012}. We define the electron temperature anisotropy as 
\begin{equation}\label{eq:anise}
    A_e = \left(\frac{2P^{(e)}_1}{P^{(e)}_2+P^{(e)}_3}\right) 
    -1,
\end{equation}
where $P_1>P_2>P_3$ are the eigenvalues derived from the  diagonalization of the pressure tensor $P_{ij}$. We show the time evolution of the rms value of $A_e$ in Fig.~\ref{fig:enet}. It reveals a marked peak at $t \sim 900 \omega_{pe}^{-1}$, the time at which the magnetic energy 
attains its maximum  production rate.

Figure~\ref{fig:3Dplots} shows a 3D representation of the turbulent pattern at $t=800\omega_{pe}^{-1}$. The (colored) filled contours represent $|\bf B|$, emerging as patterns surrounded by the vortical flow (green streamlines). The patches are localized in particular regions of the volume, i.e., the magnetic field grows in between rolling vortexes. These structures resemble the typical snake-like filaments of the Weibel instability.

\begin{figure}
\includegraphics[width=\columnwidth]{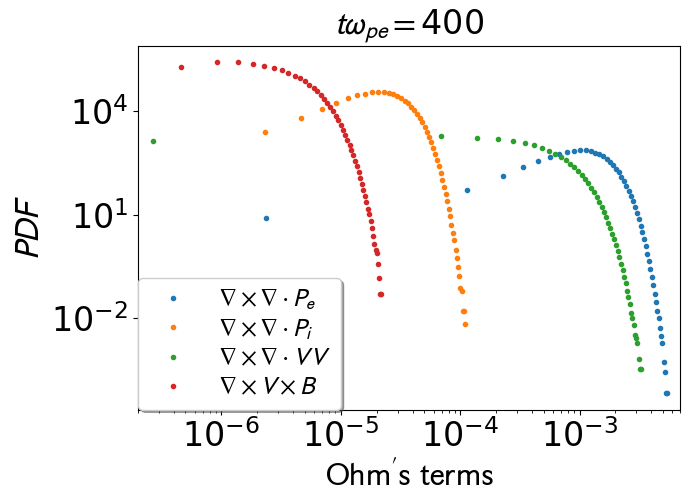}
\caption{PDF of the induction equation terms given by Eq.~(\ref{eq:Ohm}) 
at time $400\omega_{pe}^{-1}$. 
} \label{fig:OhmPDF}
\end{figure}

Starting from the induction equation, in order to understand the best candidate for the magnetic field amplification, we analyzed the separate contributions to the electric field curl coming from 
the generalized Ohm's law
\begin{equation}\label{eq:Ohm}
\begin{aligned}
   {\bf E} &\sim - \frac{1}{m_e n_i + m_i n_e}
   \left( m_e n_i {\bf v_i} + m_i n_e {\bf v_e} \right) \times {\bf B} + \\
   &\frac{m_e m_i}{e(m_e n_i + m_i n_e)} \nabla \cdot \left( n_i {\bf v_i v_i}
   - n_e {\bf v_e v_e} \right) + \\
   &\frac{m_e}{e(m_e n_i + m_i n_e)}\nabla \cdot {\mathcal {P}_i} -
    \frac{m_i}{e(m_e n_i + m_i n_e)}\nabla \cdot {\mathcal {P}_e}.
\end{aligned}
\end{equation}
Note that here we neglected contributions from terms $\propto \frac{\partial {\bm J}}{\partial t}$, as these are small compared to the other terms. We estimated the strength of the curl of each term in Eq.~(\ref{eq:Ohm}), 
by examining 
the probability density functions (PDFs) of their absolute values, as shown in Fig.~\ref{fig:OhmPDF}. It is evident that the divergence of the electron pressure tensor is the source of the magnetic field perturbations, in agreement with the behavior of the anisotropy observed in Fig.~\ref{fig:enet}.

We show the PDFs of the different terms in Fig.~\ref{fig:OhmPDF}, at time 
$t=400\omega_{pe}^{-1}$. 
The relative magnitudes of the terms do not change and we do not show here the PDFs at following times. 
The $\nabla \times \nabla \cdot {\mathcal P_e}$ term is orders of magnitude larger 
than the others, indicating that electron anisotropy 
is the leading cause of the electric field generation.

\begin{figure}
\includegraphics[width=\columnwidth]{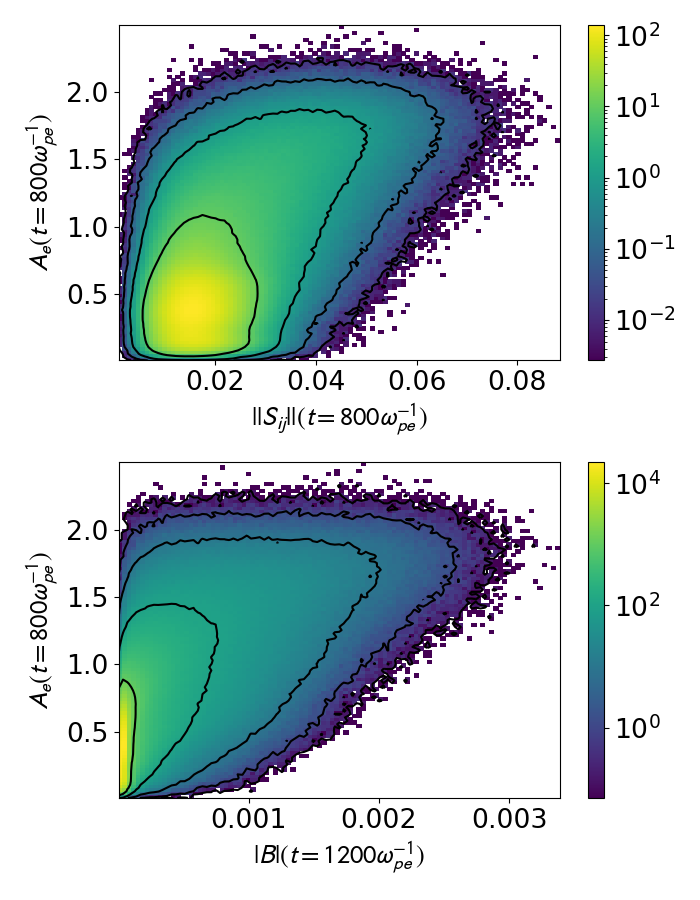}
\caption{2D histograms. Top: norm of the stress tensor and electron  anisotropy at time $t=800\omega_{pe}^{-1}$. Bottom:  magnetic field module at time $t=1200\omega_{pe}^{-1}$ and electron anisotropy at the earlier time $t=800\omega_{pe}^{-1}$. The black lines indicate the decades in the colorbars.}\label{fig:hist}
\end{figure}

We propose a mechanism for the generation of the pressure anisotropy, as follows. Integrating the Vlasov equations, one can obtain the evolution equation for the pressure tensor. In accordance with the earliest stage of the unmagnetized Taylor Green vortex, assuming for simplicity a zero heat flux closure and neglecting the electromagnetic contribution, one finds \citep{DPC2016}
\begin{eqnarray}
\frac{\partial {\mathcal P}_{i j}}{\partial t} \sim - \left[ \frac{ \partial u_k {\mathcal P}_{i j}}{\partial x_k} + 
{\mathcal P}_{k j} \frac{\partial u_i}{\partial x_k} + {\mathcal P}_{i k} \frac{\partial u_j}{\partial x_k} \right],
\label{eq:dpijdt}
\end{eqnarray}
where $u_j$ is the bulk velocity. The above approximation is true at leading order, for both species, although electrons are of central interest, since they carry currents. Assuming an initially Maxwellian distribution,  ${\mathcal P}_{i j} = P \delta_{i j}$ ($P$ being the isotropic pressure), and an incompressible flow, Eq.~(\ref{eq:dpijdt}) simplifies to
\begin{eqnarray}
\frac{\partial {\mathcal P}_{i j}}{\partial t} \sim - P \left[\frac{\partial u_i}{\partial x_j} + 
\frac{\partial u_j}{ \partial x_i} \right] \equiv -P \mathcal{S}_{i j},  
\label{eq:dpijdt2}
\end{eqnarray}
where $\mathcal{S}$ is  the stress tensor of the bulk stirring flow.
In our case, this can be obtained from Eq.~(\ref{eq:TG}). Eq.~(\ref{eq:dpijdt2}) reveals an important
property, namely that the velocity distribution function will be distorted  along the principal axes of the stress tensor;
that is, the temperature anisotropy results from the topology of the vortical Taylor Green flow.

To validate this model, we computed the 2D, joint distributions between the anisotropy and the stress tensor. 
For the stress tensor, we calculated its Frobenius norm, or second invariant, $||\mathcal{S}||=\sqrt{ Tr\{\mathcal{S} \mathcal{S}^{T}\}}$, 
where ${\rm Tr}$ indicates the trace of the matrix.
In Fig.~\ref{fig:hist} we observe that $A_e$ and $\mathcal{S}$ are correlated.  
We computed this correlation at the earliest stages, namely at $t\omega_{pe}\sim 800$, but the picture is similar at any time that precedes the anisotropy collapse in Fig.~1. Analogously, to establish the causality
of the processes, we computed the joint distributions built on $A_e$ and $|\bm B|$: at a given time, high values of the magnetic field are located in regions where the temperature anisotropy was previously large. This joint PDF is reported in Fig.~\ref{fig:hist} (bottom).

\begin{figure}
    \centering
    \includegraphics[width=0.97\columnwidth]{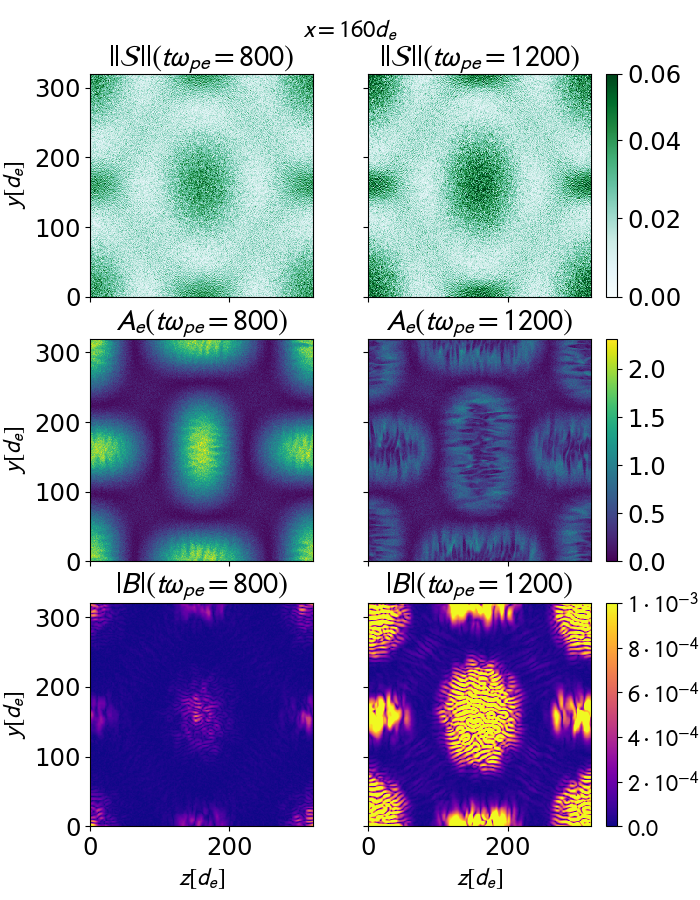}
    \caption{2D shaded contour, at the center of the box $x =10 d_i$, for electron temperature anisotropy (top), magnetic field (middle) and stress tensor strength (bottom). Left and right columns refer to $t\omega_{pe}=800$ and $1200$, respectively.
    }
    \label{fig:6panels}
\end{figure}

The above picture can be recovered locally, by looking at 2D cuts in space.  In Fig.~\ref{fig:6panels} we report $A_e$, $|\bf B|$ and $||{\mathcal S}_{ij}||$, in the middle of the domain. High $\mathcal{S}$ generates $A_e$. After the latter reaches its maximum, the electron pressure isotropizes, giving birth to the magnetic field. The missing step is now to confirm the role of the Weibel instability in this dynamics.

\begin{figure}
    \centering
    \includegraphics[width=\columnwidth]{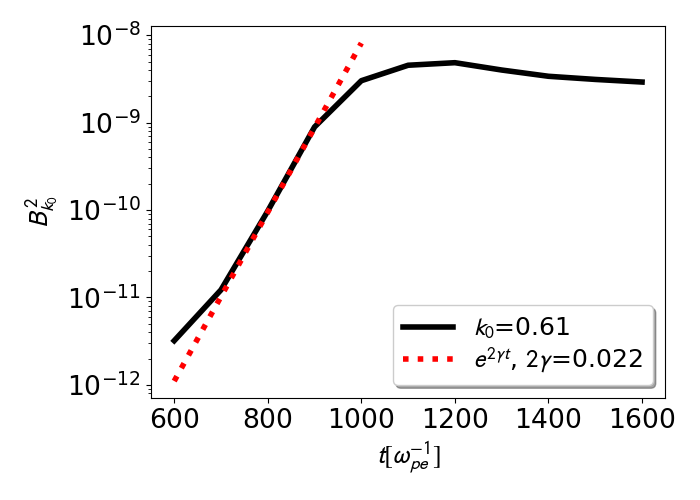}\\
    \caption{Growth rate of the Weibel instability for the fastest growing 
    mode $k_0=0.61$. 
    The red line is the fit to the exponential growth phase.}
    \label{fig:fit}
\end{figure}

We summarize the main ingredients of the electron Weibel instability \cite{Weibel59}. 
For a system where $T_{\parallel} \equiv T_{\rm 1} > T_{\perp} \equiv T_{\rm 2} \equiv T_{\rm 3}$, ${\bf k} = k\hat{\rm {\bf e}}_2$, ${\bf E} = E\hat{\rm {\bf e}}_1$ and ${\bf B} = B\hat{\rm \bf {e}}_3$, and where the initial state is defined by a bi-Maxwellian perturbed by $f = f_1 {\rm exp}\left\{ i \left( k x_2 - \omega t \right) \right\}$,
the dispersion relation becomes \cite{Krall73}
\begin{equation}\label{eq:disp}
d_e^2 k_y^2 - \frac{\omega^2}{\omega_{pe}^2} +a_e + \left( a_e+1\right) \xi \zeta(\xi) = 0.
\end{equation}
In the above, $\zeta(\xi)$ is the plasma dispersion function, $\xi = \frac{\omega}{k_y v_\perp}$, $u = \frac{v_y}{v_\perp}$, and $a_e=\frac{v_{\parallel}^2}{v_{\perp}^2}-1$. Note that the latter is equivalent to the temperature anisotropy in the minimum variance frame $A_e$, defined in Eq.~(\ref{eq:anise}). We solved Eq.~(\ref{eq:disp}) by looking for complex roots of the type $\omega = i \gamma$, using the parameters of our numerical experiment.
We show the time evolution of the fastest growing mode, which we measured as $k_0=0.61 d_e^{-1}$ from Fig.~\ref{fig:enet} (bottom). In Fig.~\ref{fig:fit} we report the time history of this mode, which grows exponentially in time and then saturates. From the exponential fit we get $\gamma=0.011$. The Weibel theoretical growth rate is obtained using $v_\perp=v_{th,e}$ and $a_e=2$, the latter being close to the maximum value shown in Figures \ref{fig:hist} and \ref{fig:6panels}. These parameters give $\gamma_{th}=0.015$, in agreement with our numerical result.

\section{Conclusions}

In this Letter, we demonstrated that a magnetic field can be generated by shearing motions typical of turbulent fields, via an electron-Weibel instability. Our initial condition is meant to mimic any vortical motion, such as the eddy interaction in convective cells or the turbulence that may develop in coherent shearing motions triggered by gravitational perturbations \cite{ParievColgate2007I,ParievColgate2007II}. The stresses generated at the intersection of the rotating vortexes produce pressure anisotropy,  qualitatively in accord with Eq.s~(\ref{eq:dpijdt})-(\ref{eq:dpijdt2}). The ions, being heavier, are slow and play very little role in the early system dynamics. On the other hand, the lighter electrons interact faster with the fields, generating a Weibel instability that, in turn, induces exponential growth of the magnetic field. 
This hypothesis is supported by our analysis of Ohm's law, where the divergence of the electrons pressure is the dominant term. As shown in Figure~\ref{fig:6panels}, the electron anisotropy at two consecutive times shows the imprint of the magnetic field generation. The theoretical growth rate matches well with the most energetic mode emerging the simulation. Such mechanism provides a plausible explanation for the origin of the seed magnetic field necessary for dynamo theories. 

At later times, by looking at the magnetic spectra, the magnetic energy is back-transferred to larger scales, probably due to a complex, nonlinear interaction with the cascading bulk flows. This interaction may represent the onset of a nonlinear regime of a fully kinetic dynamo action, a topic to be be investigated in future works.
\begin{acknowledgments}
This work has received funding from the European
Union’s Horizon 2020 Research and Innovation Program Grant 776262
(Artificial Intelligence Data Analysis; http://www.aida-space.eu/). F.P. is supported by the PostDoctoral Fellowship 12X0319N and the Research Grant 1507820N from Research Foundation -- Flanders (FWO).
We acknowledge the European PRACE initiative for awarding us access to the supercomputer SuperMUC-NG at GCS@LRZ, Germany.
\end{acknowledgments}

%

\bibliography{paper}{}
\bibliographystyle{aasjournal}



\end{document}